\documentclass[letterpaper, 10 pt, conference]{ieeeconf} 
\usepackage{CJKutf8,amsmath,graphicx}
\usepackage{multirow}
\usepackage{array}
\usepackage{cite}
\usepackage{amsmath,amssymb,amsfonts}
\usepackage{graphicx}
\usepackage{textcomp}
\usepackage{multirow}
\usepackage{array}
\usepackage{tikz}
\usepackage{footnote}
\usepackage[ruled,vlined, linesnumbered]{algorithm2e}

\IEEEoverridecommandlockouts                              

\title{\LARGE \bf
Decoding Neural Correlation of Language-Specific Imagined Speech using EEG Signals
}

\author{Keon-Woo Lee$^{1}$, Dae-Hyeok Lee$^{1}$, Sung-Jin Kim$^{2}$, and Seong-Whan Lee$^{2}$%
,~\IEEEmembership{Fellow,~IEEE}
\thanks{\copyright 20XX IEEE.  Personal use of this material is permitted. Permission from IEEE must be obtained for all other uses, in any current or future media, including reprinting/republishing this material for advertising or promotional purposes, creating new collective works, for resale or redistribution to servers or lists, or reuse of any copyrighted component of this work in other works.}
\thanks{This research was supported by `Defense Challengeable Future Technology Program' of Agency for Defense Development, Republic of Korea.}
\thanks{$^{1}$Keon-Woo Lee and Dae-Hyeok Lee are with Department of Brain and Cognitive Engineering, Korea University, 145, Anam-ro, Seongbuk-gu, Seoul 02841, Republic of Korea 
        {\tt\small \{kw-lee and lee\_dh\}@korea.ac.kr}}%
\thanks{$^{2}$Sung-Jin Kim and Seong-Whan Lee are with Department of Artificial Intelligence, Korea University, 145, Anam-ro, Seongbuk-gu, Seoul 02841, Republic of Korea 
        {\tt\small \{s\_j\_kim and sw.lee\}@korea.ac.kr}}%
}

\begin{document}

\maketitle
\thispagestyle{empty}
\pagestyle{empty}

\begin{abstract}

Speech impairments due to cerebral lesions and degenerative disorders can be devastating. For humans with severe speech deficits, imagined speech in the brain--computer interface has been a promising hope for reconstructing the neural signals of speech production. However, studies in the EEG--based imagined speech domain still have some limitations due to high variability in spatial and temporal information and low signal--to--noise ratio. In this paper, we investigated the neural signals for two groups of native speakers with two tasks with different languages, English and Chinese. Our assumption was that English, a non-tonal and phonogram--based language, would have spectral differences in neural computation compared to Chinese, a tonal and ideogram--based language. The results showed the significant difference in the relative power spectral density between English and Chinese in specific frequency band groups. Also, the spatial evaluation of Chinese native speakers in the theta band was distinctive during the imagination task. Hence, this paper would suggest the key spectral and spatial information of word imagination with specialized language while decoding the neural signals of speech. 
\newline

\indent \textit{Clinical relevance}— Imagined speech--related studies lead to the development of assistive communication technology, especially for patients with speech disorders such as aphasia due to brain damage. This study suggests significant spectral features by analyzing cross--language differences of EEG--based imagined speech using two widely used languages.
\end{abstract}

\section{INTRODUCTION}

Brain-computer interface (BCI) has attracted much interest in biomedical engineering fields in pursuit of rehabilitation matters, such as robotic arms and spelling system\cite{c1, c2, c3}. In many BCI applications, imagined speech, which is defined as an internal pronunciation of words without any audible output, has also been studied for neural rehabilitation. It has been immersed by many professionals to assist people with disabilities such as aphasia or locked-in syndrome due to amyotrophic lateral sclerosis \cite{c4, c5, c6, c7}. Noticeable achievements of imagined speech--based studies have been brought using various signal processing approaches. Especially, the studies using an invasive method electrocorticography (ECoG) have contributed the significant improvement in decoding signals of imagined speech with a natural communication rate \cite{c8, c9}. 

To accomplish the fundamental goal of imagined speech, electroencephalogram (EEG) studies also must be continuously advanced in parallel with the studies using ECoG and other imaging methods, since the broad variety of data and diverse aspects of observation are in demand. Imagined speech studies using EEG signals have attempted to utilize various computer--based approaches including machine learning algorithms to calibrate and interpret the neural signals \cite{c10}. EEG is one of the most effective brain imaging techniques since it is non-invasive, easy-to-use, and highly applicable in a real-world environment\cite{c11, c12, c13}.

However, Proix \textit{et al.} \cite{c14} referred that the studies in imagined speech still have some limitations because of the low signal--to--noise ratio in raw data. Moreover, the neural signals of imagined speech have several challenges to carry out; \textit{i}) clear findings of neural pathways during imagination of speech production, \textit{ii}) reduction of variability in spatial and temporal decoding, and \textit{iii}) perfect exclusion of confounding factors in experiments.


In order to fully understand the language processing of imagined speech using EEG signals, understanding the neural representation of various languages is essential. Speech production requires a strong interaction between cognitive process and articulation, and language is one of the key components for lexical, syntactic, and semantic speech processing. Hence, in this paper, we investigate the neural correlates of language-specific imagined speech, which is based on the native language (NL) for each person and the language for the imagination task (LT). Namely, our alternative hypothesis is that the language-specific imagined speech must be decoded in discriminative neural computation. We selected two widely used languages, English and Chinese. 
English is characterized as a non-tonal and phonogram--based language, which implies that the tonal differences do not alter the meaning of the words and each letter does not contain any meaning itself \cite{c15}. In contrast, Chinese syllables have four major tonal difference, and each tonal differences have a specific meaning of its word, such as the flat tone, the rising tone, the falling-rising tone, and the falling tone \cite{c16, c17}. Moreover, Chinese words are built as ideogram--based language, which means that each letter contains a meaning. We observed spectral power differences of neural signals for two LTs with two groups of NL speakers. We could observe the significant difference by analyzing the power distribution and scalp topographic results using EEG signals. This observation has been the first attempt in the imagined speech-EEG domain to the best of our knowledge.

\section{MATERIALS AND METHODS}

\subsection{Participants}

We acquired EEG signals from 12 healthy participants (two males and 10 females; age 27.5±1.73 years). All of them were confirmed as right-handed. The participants were gathered and divided into two NL groups as six English native speakers (EN) and six Chinese native speakers (CN). This study was approved by Institutional Review Board at Korea University (KUIRB-2020-0319-01), and written informed consent was obtained from all participants prior to the experiment.

\subsection{Experimental Protocols}

The experiment was designed into two sessions with two different languages, English tasks (ET) and Chinese tasks (CT), respectively. Both ET and CT used identical paradigms as shown in Fig. 1(a). Prior to the non-NL tasks, the quiz for four words was given to all of the participants in order to ensure that they were familiar with the foreign language. 

As shown in Fig. 1(a), each trial consists of a visual cue, beep sounds, `+' (preparation for imagination), four imagination times with blank screens, and bold `+' (rest). Between the visual cue-imagination and imagination-rest, the beep sounds were presented to notify the participants of either the preparation of imagination or the end of imagination. For the imagination times, the blank screens were presented in a regular sequence and the participants were instructed to imagine the specific word without any movements of muscles for speech production. 

\begin{CJK*}{UTF8}{gbsn}
The words for imagination were selected based on the application of drone swarm control in BCI \cite{c18}. The meaning of words for ET and CT were equivalently paired, for instance, ``Go" and ``走~[Zǒu]", meaning ``Go" in Chinese. We also instructed the participants to imagine `speaking' the words and not the scene of a moving drone. The chosen words for ET were ``Go", ``Follow Me", ``Return", and ``Stop". The words for CT with identical meanings were ``走~[Zǒu]", ``跟着我~[Gēnzhe wǒ]", ``返回~[Fǎnhuí]", and ``停止~[Tíngzhǐ]", respectively.

\end{CJK*}

\subsection{Data Acquisition and Preprocessing}
All participants were seated on a comfortable chair, and the partition was used to minimize the disturbance during the experiment as shown in Fig. 1(b). EEG signals for this experiment were acquired at 1,000 Hz using 64 electrodes according to the international 10-20 system via BrainVision recorder (BrainAmp, Brain Products GmBH, Germany) as shown in Fig. 1(c). The ground and reference electrodes were FPz and FCz channels, respectively. All electrodes were set at impedance levels below 10 k$\Omega$, and a notch filter of 60 Hz was applied to reduce power supply noise. Also, the band-pass filter was used between 0.1 and 125 Hz. Then, the independent component analysis was used to remove the artifacts such as eye blinks or muscle movements.

The subgroups of datasets were divided into four groups based on the NLs and LTs. We named each subgroup, and the four subgroups: `EN/ET', `EN/CT', `CN/ET', and `CN/CT'. 

The five frequency band groups were the delta (0.1--4 Hz), theta (4--8 Hz), alpha (8--12 Hz), beta (12--30 Hz), and gamma (30--125 Hz) bands, which have been widely used in the analysis of EEG signals \cite{c7}. 


\begin{figure}[t!]
\centering
\scriptsize
\centerline{\includegraphics[width=0.85\columnwidth, height=0.30\textwidth]{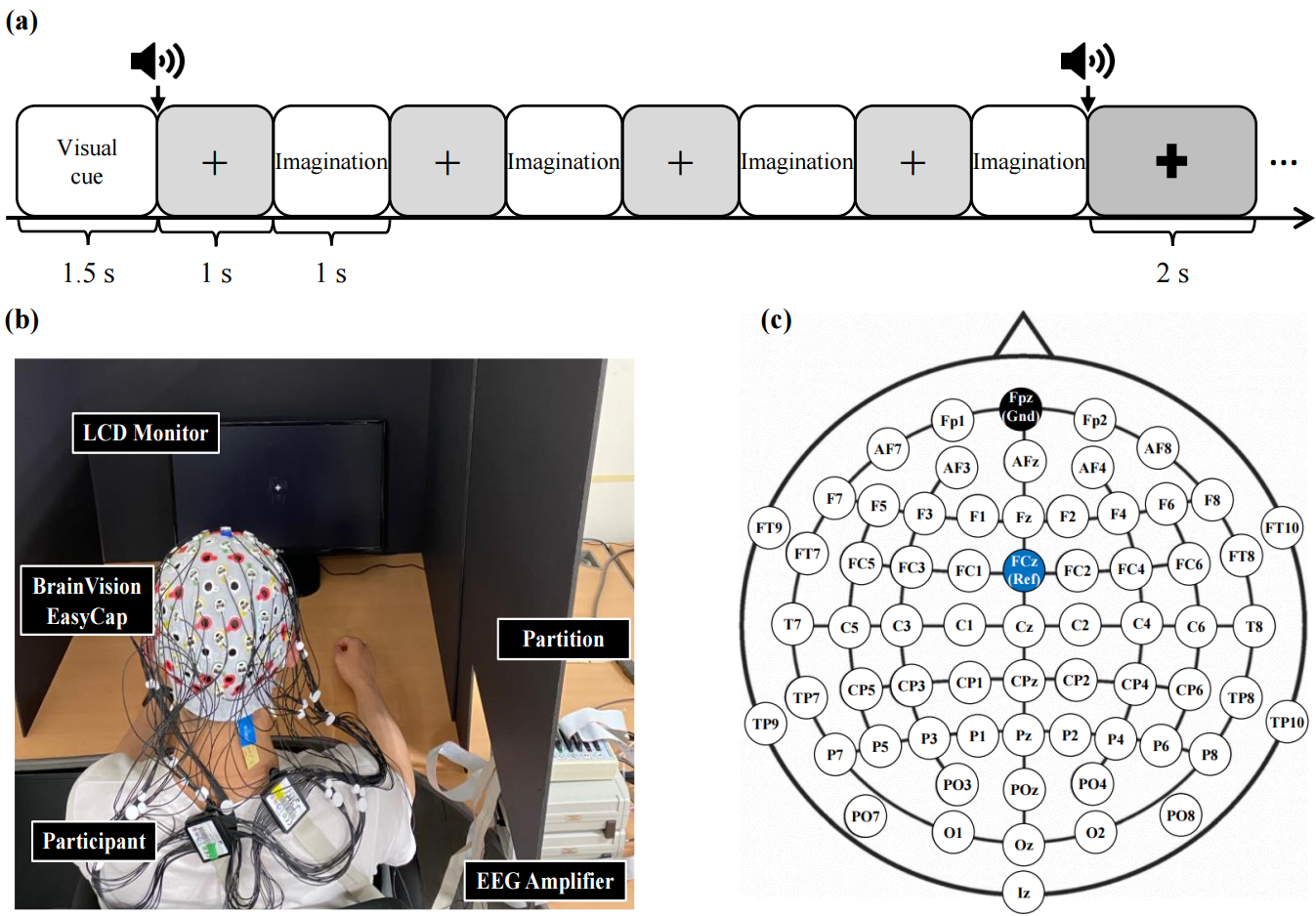}}
\caption{Description of the experimental setup and channel locations. a) Experimental paradigm for both English and Chinese tasks. Each trial consists of four imagination times for the presented words as a visual cue. b) Experimental environment for imagined speech tasks. c) Channel locations for 64 electrodes.}
\end{figure}

\subsection{Relative Power Spectral Density}

The power spectral density (PSD) is widely used to evaluate the EEG signals, especially because it describes the power distribution of EEG series in frequency domain \cite{c19}. Based on the trigger information of imagination segments, the PSD was measured for each frequency: 
\begin{equation}
{PSD_{f_2-f_1}} = 10*\log_{10}(2\int_{f_1}^{f_2}|\hat{x}(2{\pi}f)|^2 df) 
\end{equation}
where $f_1$ was lower frequency and $f_2$ was higher frequency. $\hat{x}(2{\pi}f)$ was measured during the fast Fourier transform calculation \cite{c20}.\\

We used the relative PSD (RPSD) in order to fairly compare the power of signals in cross-subject \cite{c21, c22}. 
\begin{equation}
{P_{relative}} = \frac{\sum^{f_2}_{f=f_1}P(f)}{\sum^{f_H}_{f=f_L}P(f)}
\end{equation}
As shown in Eq. 2, the RPSD was obtained by dividing the PSD of targeted bands by the total PSD of whole bands.$f_1$ and $f_2$ represent the targeted band groups (e.g., $f_1=4$ and $f_2=8$ when the theta band is chosen), and $f_L=0.1$ and $f_H=125$, which fixed as the total PSD. The RPSD could offer information on the proportion in each band group and the significance. Moreover, in order to test the neural correlation between different combinations of subgroups, we subtracted the RPSD with six possible subtractions. The statistical significance using the significance level below 0.05 was performed for all RPSD. 

\begin{figure}[t]
\centering
\scriptsize
\centerline{\includegraphics[width = 0.85\columnwidth, height=0.27\textheight]{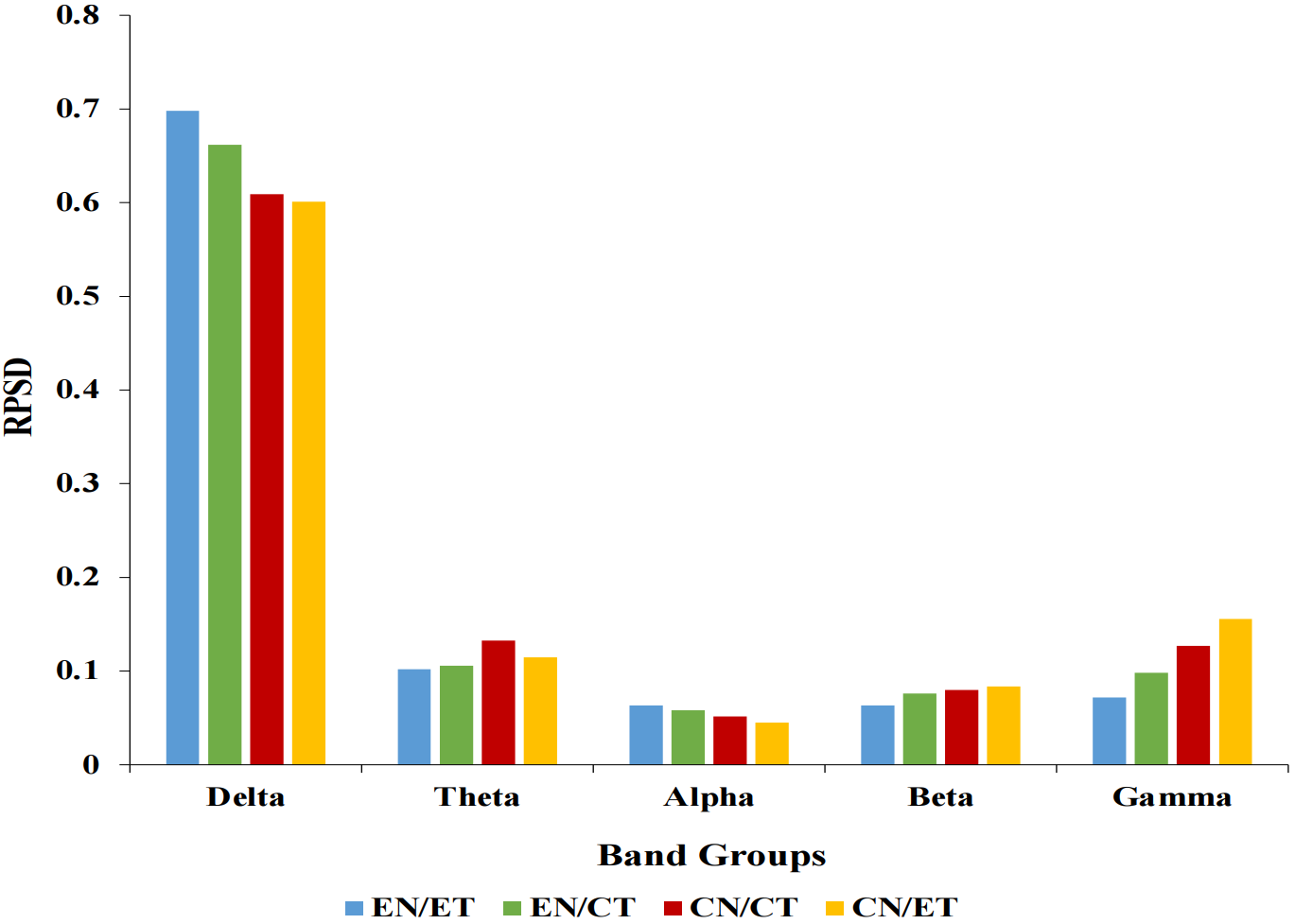}}
\caption{The bar plots of the averaged RPSD across participants, trials, and channels for the band groups. The four subgroups were divided into the five band groups, and the RPSD was represented by the bar plots.}
\end{figure}

\begin{figure}[t]
\centering
\scriptsize
\centerline{\includegraphics[width=0.75\columnwidth, height=0.24\textheight]{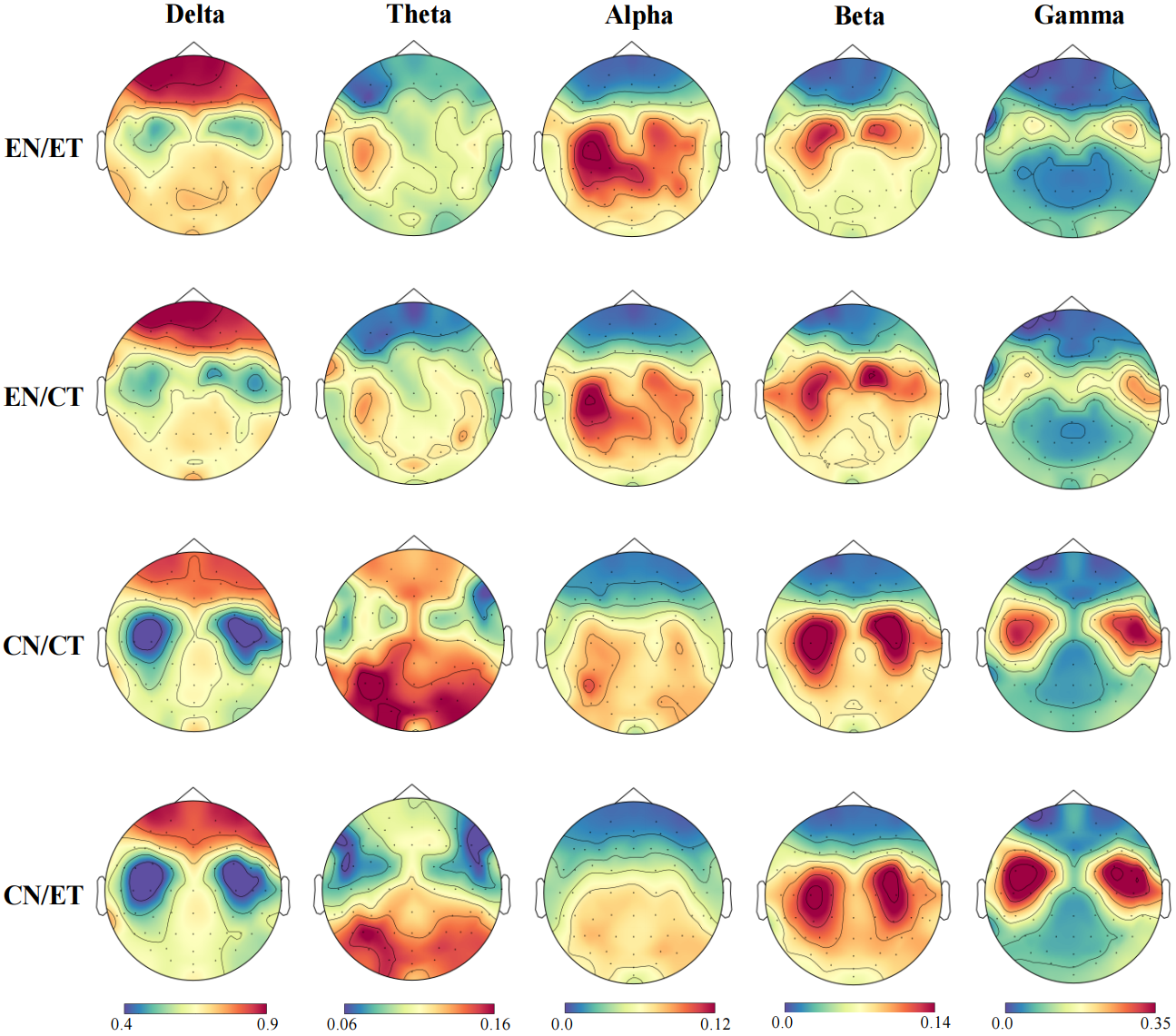}}
\caption{The channel-wise RPSD for the subgroups in scalp topographies. Each scalp topography indicates the RPSD within the subgroups for the five different band groups. The scale bars below the topographies represent the scale used for the topographies in each band group.}
\end{figure}

\begin{figure}[t]
\centering
\scriptsize
\centerline{\includegraphics[width=0.75\columnwidth, height=0.24\textheight]{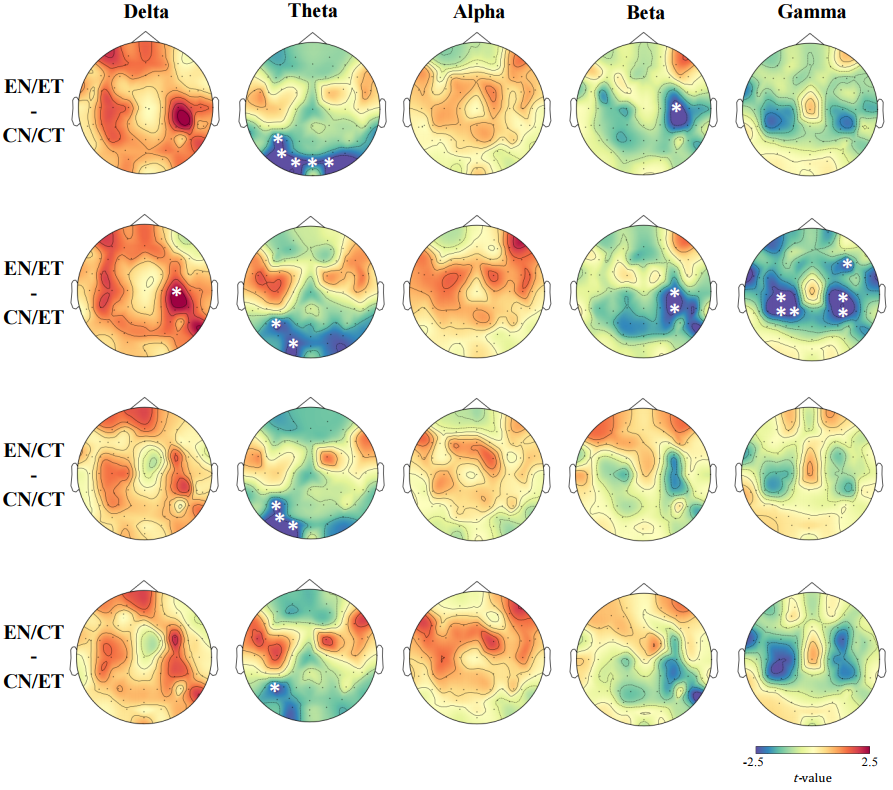}}
\caption{Inter--NL differences of RPSD between subgroups in \textit{t}--value scalp topographies. Each scalp topography indicates the distribution of channels with \textit{t}--value using the paired $\textit{t}$‐test in the significance level below 0.05. The locations of channels with the statistical significance are indicated as white `$\ast$' (*: \textit{p}$<$0.05).}
\end{figure}

\begin{figure}[t]
\centering
\scriptsize
\centerline{\includegraphics[width=0.75\columnwidth, height=0.13\textheight]{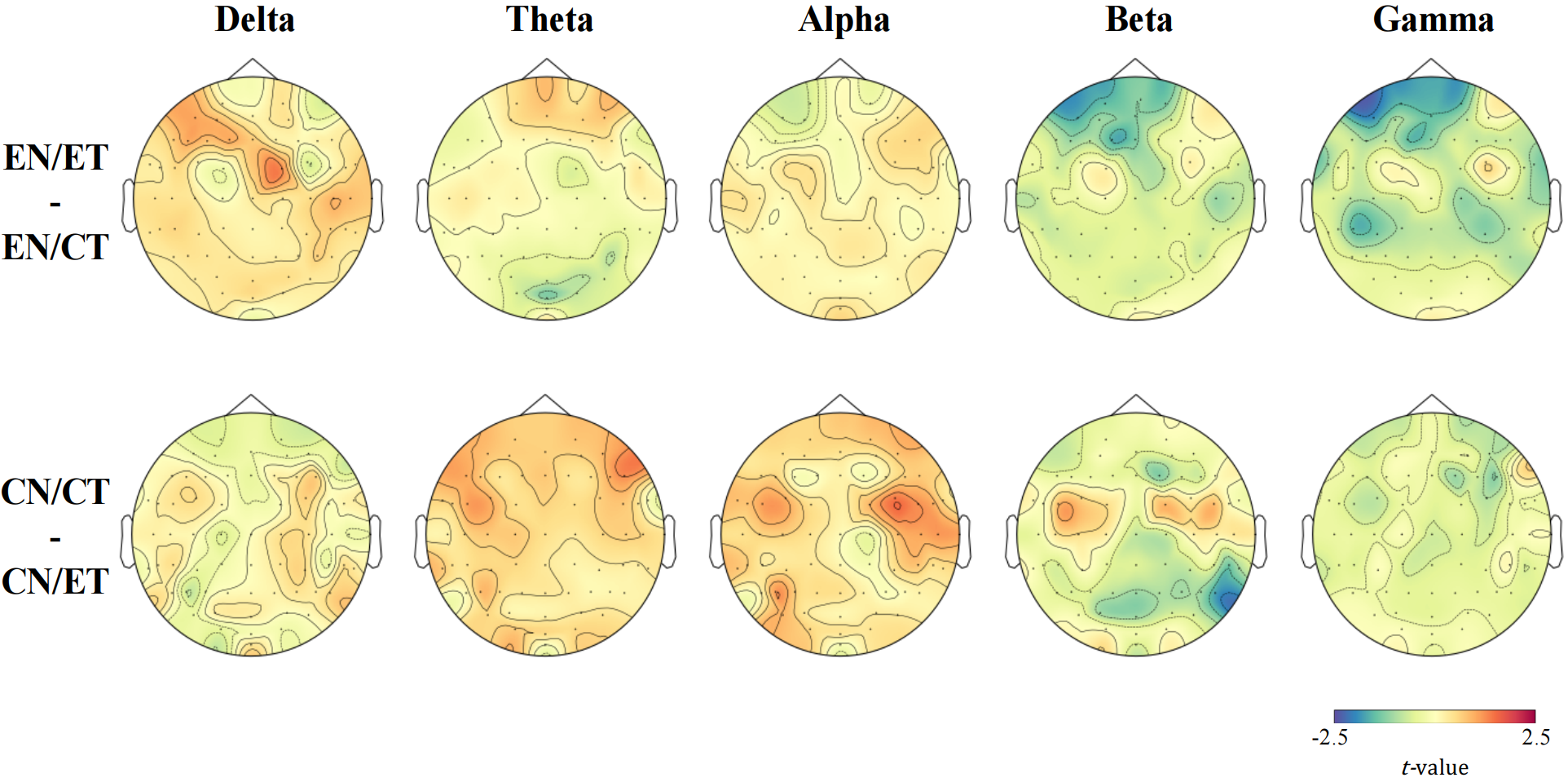}}
\caption{Intra--NL differences of RPSD between subgroups in \textit{t}--value scalp topographies. No statistically significant difference for two language tasks within the same NL groups was observed.}
\end{figure}


\section{RESULTS}
Fig. 2 shows the bar plots of the averaged RPSD for each subgroup in the five different band groups. All subgroups had the highest RPSD in the delta band. 


\subsection{Analysis of RPSD for each subgroup}
As shown in Fig. 3, the RPSD of four subgroups in the five band groups were interpolated into the scalp topographies. All of them had high distribution in the prefrontal cortex for the delta band groups. In the theta band group, both CN groups had a higher proportion of power in the central parietal and occipital lobes. Also, for the beta and gamma band groups, CN had stronger activation in the motor and somatosensory cortex. In contrast, EN groups had the higher RPSD in the central and temporal lobes for the alpha band. Since the suppression of the alpha band is correlated to the discrimination of visual target, CN groups might have a higher ability to differentiate the imagination process for four different words \cite{c23}. 

\subsection{RPSD differences for NL comparisons} 
We used the paired \textit{t}‐test of RPSD difference in each channel location using the significance level below 0.05 in order to observe the distinctive PSD feature in combination with spatial information \cite{c7}. The significant difference of RPSD in each channel location based on the subgroups and the band groups was successfully observed using calculation of the statistical significance and interpolation of \textit{t}--values into the topographies. 

The RPSD differences in topographies can be divided into two analyses, which are inter-- and intra--NL comparisons. Inter--NL analysis is to observe the difference between different NL groups for two LTs. As shown in Fig. 4, plotted topographies for inter--NL comparisons were obtained by calculation of \textit{t}--values using subtraction of RPSD such as EN/ET--CN/CT, EN/ET--CN/ET, EN/CT--CN/CT, and EN/CT--CN/ET, respectively. This enabled us to compare the distribution of power across conditions with varying detail. 

For the delta band group, both EN/ET subgroups had higher activation on the right hemisphere of the central lobe, especially the significant difference was observed between EN/ET and CN/ET on C4 channel. This result could also be expected based on Fig. 3, since CN groups had the significantly lower power in the motor and somatosensory cortex for the delta band. In contrast, CN groups had stronger activation in the same areas during the beta and gamma bands, which could be observed in both Fig. 3 and 4. Since high activation of the sensorimotor cortex for tonal differences was studied earlier, observation in Fig. 4 was in line with the past studies \cite{c17}. However, the significant difference was produced only when CN groups are compared EN/ET group. 

The most important finding was that all inter--native comparisons had the significant difference in the inferior parietal lobe including a wide range of central-parietal or occipital lobe during the theta band    . Especially, although both groups used their NL, a stronger difference was made between EN/ET and CN/CT on O1, Oz, O2, P5, and PO7 channels. Most of these channels locate on the inferior parietal lobe, which is a well-known area as a multimodal brain region related to spatial attention, language, and higher motor processing \cite{c24}. Hence, CN might have stronger attention to language and higher motor processing based on the topographies. 

Intra--NL comparisons were also made by subtracting the RPSD for identical NL speakers with different LT. The plotted \textit{t}--values in topographies are negligible based on Fig. 5. Hence, the language processing for the same NL speakers with a different language was hard to discriminate based on this EEG analysis.


\section{DISCUSSION AND CONCLUSION}

We have several limitations in this paper. First, although the imagination tasks in our experiment were built to classify the words using machine learning algorithms, we did not evaluate the classification performance at this time. However, the goal of this study was to observe the distinction of neural signals for different languages, therefore the classification performance might not be the key evaluation of the index. Moreover, by applying the language-specific features that we observed in this work, we will develop the algorithm for classification eventually. Also, the absolute value of PSD was not discussed in this work. Although we obtained all PSD values for subgroups and the band groups, we did not use them to evaluate at this time since the PSD can be highly varied among the participants and the RPSD can minimize the participant dependent variance, especially for the subtraction \cite{c22}.

Our major goal of this study was to analyze the neural correlation of imagined speech in different languages. By comparing the RPSD in the combinations of NL and LT, we were able to find that the spectral feature and the distribution of power had the significant distinction for different NL speakers. Hence, we found that the difference in spectral features according to language characteristics is an important factor when decoding imagined speech-based EEG signals. 

\section*{ACKNOWLEDGMENT}

The authors would like to thank H.-J. Ahn for the support extended to this work.


\begin{thebibliography}{99}

\bibitem{c1} J.-H. Jeong, K.-H. Shim, D.-J. Kim, and S.-W. Lee, "Brain-controlled robotic arm system based on multi--directional CNN--BiLSTM network using EEG signals," \textit{IEEE Trans. Neural Syst. Rehabil. Eng.}, vol. 28, no. 5, pp. 1226--1238, 2020.

\bibitem{c2} D.-O. Won, H.-J. Hwang, D.-M. Kim, K.-R. Müller, and S.-W. Lee, "Motion--based rapid serial visual presentation for gaze--independent brain--computer interfaces," \textit{IEEE Trans. Neural Syst. Rehabil. Eng.}, vol. 26, no. 2, pp. 334--343, 2017.

\bibitem{c3} M.-H. Lee, J. Williamson, D.-O. Won, S. Fazli, and S.-W. Lee, "A high performance spelling system based on EEG--EOG signals with visual feedback," \textit{IEEE Trans. Neural Syst. Rehabil. Eng.}, vol. 26, no. 7, pp. 1443--1459, 2018.

\bibitem{c4} C. Cooney, R. Folli, and D. Coyle, “Neurolinguistics research advancing development of a direct--speech brain--computer interface,” \textit{IScience}, vol. 8, pp. 103–125, 2018.

\bibitem{c5} N. Kaongoen, J. Choi, and S. Jo, “Speech--imagery--based brain--computer interface system using ear--EEG,” \textit{J. Neural Eng.}, vol. 18, no. 1, p. 016023, 2021.

\bibitem{c6} G. Krishna, C. Tran, Y. Han, M. Carnahan, and A. H. Tewfik, “Speech synthesis using EEG,” in \textit{IEEE Int. Conf. Acoust. Speech Sig. Proc. (ICASSP)}, Barcelona, Spain, 2020, pp. 1235–1238.

\bibitem{c7} S.-H. Lee, M. Lee, and S.-W. Lee, “Neural decoding of imagined speech and visual imagery as intuitive paradigms for BCI communication,” \textit{IEEE Trans. Neural Syst. Rehabil. Eng.}, vol. 28, no. 12, pp. 2647–2659, 2020.

\bibitem{c8} G. K. Anumanchipalli, J. Chartier, and E. F. Chang, “Speech synthesis from neural decoding of spoken sentences,” \textit{Nature}, vol. 568, no. 7753, pp. 493–498, 2019.

\bibitem{c9} S. Martin, I. Iturrate, J. R. Mill{\'a}n, R. T. Knight, and B. N. Pasley, “Decoding inner speech using electrocorticography: Progress and challenges toward a speech prosthesis,” \textit{Front. Neurosci.}, vol. 12, p. 422, 2018.

\bibitem{c10} P. Saha, S. Fels, and M. Abdul-Mageed, “Deep learning the EEG manifold for phonological categorization from active thoughts,” in \textit{IEEE Int. Conf. Acoust. Speech Sig. Proc. (ICASSP)}, Brighton, UK, 2019, pp. 2762–2766.

\bibitem{c11} O.-Y. Kwon, M.-H. Lee, C. Guan, and S.-W. Lee, "Subject--independent brain--computer interfaces based on deep convolutional neural networks," \textit{IEEE Trans. Neural Netw. Learn. Syst.}, vol. 31, no. 10, pp. 3839--3852, 2020.

\bibitem{c12} C. H. Nguyen, G. K. Karavas, and P. Artemiadis, “Inferring imagined speech using EEG signals: A new approach using Riemannian manifold features,” \textit{J. Neural Eng.}, vol. 15, no. 1, p. 016002, 2017.

\bibitem{c13} M.-H. Lee \textit{et al.}, "EEG dataset and OpenBMI toolbox for three BCI paradigms: An investigation into BCI illiteracy," \textit{GigaScience}, vol. 8, no. 5, pp. 1--16, 2019.

\bibitem{c14} T. Proix \textit{et al.}, “Imagined speech can be decoded from low--and cross--frequency intracranial EEG features,” \textit{Nat. Commun.}, vol. 13, no. 1, pp. 1–14, 2022.

\bibitem{c15} T. Nabeshima and Y.-P. Gunji, “Zipf’s law in phonograms and weibull distribution in ideograms: Comparison of English with Japanese,” \textit{BioSystems}, vol. 73, no. 2, pp. 131–139, 2004.

\bibitem{c16} Y. Li, C. Tang, J. Lu, J. Wu, and E. F. Chang, “Human cortical encoding of pitch in tonal and non--tonal languages,” \textit{Nat. commun.}, vol. 12, no. 1, pp. 1–12, 2021. 

\bibitem{c17} X. Zhang, H. Li, and F. Chen, “EEG--based classification of imaginary mandarin tones,” in \textit{Int. Conf. IEEE Eng. Med. Biol. Soc. (EMBC)}, Montreal, Canada, 2020, pp. 3889–3892.

\bibitem{c18} P. Peining, G. Tan, and A. A. P. Wai, “Evaluation of consumer--grade EEG headsets for BCI drone control,” in \textit{IRC Conf. Sci. Eng. Tech. (IRC-SET)}, Singapore, 2017, pp. 1–6.

\bibitem{c19} H. H. Bulthoff, S.-W. Lee, T. Poggio, and C. Wallraven, \textit{Biologically motivated computer vision}. Springer, Verlag, 2003.

\bibitem{c20} G.-H. Shin, M. Lee, H.-J. Kim, and S.-W. Lee, “Prediction of event related potential speller performance using resting--state EEG,” in \textit{Int. Conf. IEEE Eng. Med. Biol. Soc. (EMBC)}, Montreal, Canada, 2020, pp. 2973–2976.

\bibitem{c21} R. Wang \textit{et al.}, “Power spectral density and coherence analysis of Alzheimer’s EEG,” \textit{Cogn. Neurodyn.}, vol. 9, no. 3, pp. 291–304, 2015.

\bibitem{c22} M. A. Rahman \textit{et al.}, “Emotion recognition from EEG--based relative power spectral topography using convolutional neural network,” \textit{Array}, p. 100072, 2021.

\bibitem{c23} H. Van Dijk, J.-M. Schoffelen, R. Oostenveld, and O. Jensen, “Prestimulus oscillatory activity in the alpha band predicts visual discrimination ability,” \textit{J. Neurosci.}, vol. 28, no. 8, pp. 1816–1823, 2008.

\bibitem{c24} S. Caspers \textit{et al.}, “Probabilistic fibre tract analysis of cytoarchitectonically defined human inferior parietal lobule areas reveals similarities to macaques,” \textit{Neuroimage}, vol. 58, no. 2, pp. 362–380, 2011.



\end{thebibliography}
\end{document}